\documentclass{PoS}

\title{Probing Gluon Helicity with Dijets from $\sqrt s$ = 510 GeV Polarized Proton Collisions at STAR}

\ShortTitle{Dijet $A_{LL}$}

\author{\speaker{Suvarna Ramachandran for the STAR Collaboration}\thanks{Heartfelt gratitude to Dr. Renee Fatemi for her guidance and support.}\\
        University of Kentucky\\
        E-mail: \email{suvarna.r@uky.edu}}

\abstract{The production of jets in polarized proton collisions at STAR is dominated by quark-gluon and gluon-gluon scattering processes. The dijet longitudinal double-spin asymmetry $(A_{LL})$ is sensitive to the helicity distributions and may be used to extract information about the gluon helicity contribution  $\Delta{g(x,Q^2)}$ to the 
spin of the proton. Previous STAR jet measurements at $\sqrt s$ = 200 GeV show evidence of polarized gluons 
for gluon momentum fractions above 0.05. The measurement of dijet $A_{LL}$ at $\sqrt s$ = 510 GeV will extend 
the current constraints on  $\Delta{g(x,Q^2)}$ to lower gluon momentum fractions and allow for the reconstruction of the 
partonic kinematics at leading order. These proceedings present preliminary results from the dijet $A_{LL}$  
measurement from $\sim50 pb^{-1}$ of  $\sim50 \%$ polarized proton data taken during the 2012 RHIC run.}

\FullConference{XXIV International Workshop on Deep-Inelastic Scattering and Related Subjects\\
		11-15 April, 2016\\
		DESY Hamburg, Germany}

\begin{document}
\section{Introduction}

The proton is a complex subatomic particle consisting of the elementary particles, quarks and gluons (collectively
known as partons). Protons are spin $\frac{1}{2}$ fermions and a key question is how the proton's spin is 
distributed amongst its constituents. The intrinsic spin of the proton can thus be expressed as:
\begin{equation}
\frac{1}{2} = \frac{1}{2}\Delta \Sigma + \Delta G + L_{Q} + L_{G}
\end{equation}
as formulated by Jaffe and Manohar ~\cite{Jaffe:1989jz}, where $\frac{1}{2}\Delta \Sigma$ and $\Delta G$ are the spin
contributions from the quarks and gluons and $L_{Q}
$ and $L_{G}$ are the orbital angular momentum contributions.  We now know, from polarized deep inelastic 
scattering experiments with leptons and protons, that the quark spin contribution is $\sim25\%$~\cite{deFlorian:2014yva}~
\cite{Nocera:2014gqa}. These same experiments provided very little information about $\Delta G$.
	
Because gluons do not carry electric charge, polarized deep inelastic scattering experiments are only sensitive to the gluon at next-to-leading order. In contrast, polarized proton-proton collisions provide access to the gluon helicity distribution, $\Delta{g(x,Q^2)}$, at leading order. The 
Relativistic Heavy Ion Collider (RHIC)~\cite{Alekseev:2003sk} at Brookhaven National Laboratory is the world's first and only polarized proton 
collider. The spin program at RHIC aims to probe $\Delta{g(x,Q^2)}$  via the measurement of the longitudinal spin asymmetry $A_{LL}$ of a range of observables, including inclusive jets and pions. $A_{LL}$ is defined as:
\begin{equation}
A_{LL} = \frac{\sigma^{++}-\sigma^{+-}}{\sigma^{++}+\sigma^{+-}}
\end{equation}
where $\sigma^{++}$ and $\sigma^{+-}$ are the cross-sections for the final states with the helicities of the protons 
aligned and anti-aligned, respectively. The 2009 STAR inclusive jet results~\cite{Adamczyk:2014ozi} at $\sqrt s$ = 200 GeV provided the
first experimental evidence of a non-zero $\Delta{g(x,Q^2)}$ in the range $x>0.05$, where $x$ is the momentum fraction of the parton inside the proton. But the lower $x$ regions are still highly
unconstrained, contributing to the large uncertainty on the total integral $\Delta{G}$. To expand the kinematic coverage and reduce the uncertainity at lower $x$, these measurements
are being extended to $\sqrt s$ = 510 GeV. 

To date, measurements have been focused on inclusive channels, which provide the smallest statistical errors.  It is not possible however to extract $x_1$ and $x_2$ of the partons participating in the interaction on an event-by-event level,  limiting the ability of these measurements to constrain the functional form of $\Delta{g(x,Q^2)}$. Correlation observables, such as dijets, provide access to the parton-level kinematics at leading order. This contribution will discuss the dijet longitudinal double spin asymmetry $A_{LL}$, for the polarized proton data at 510 GeV from 2012. The average beam polarization recorded was about 50\% and the integrated luminosity was about 50 $pb^{-1}$.

\section{RHIC and STAR Detector}
 
Polarized hydrogen beams are produced at the Optically Pumped Polarized Ion Source, and are accelerated by a Radio Frequency Quadrupole magnet and 200 MHz linear accelerator to 200 MeV. Then the electrons are stripped off and the polarized protons are injected into the Booster to be accelerated up to 2.5 GeV, before being injected into the Alternating Gradient Synchrotron where they are accelerated up to 24 GeV. They are then injected into the RHIC rings where they
can be accelerated up to various center of mass energies ranging from 62 GeV to 510 GeV. When injected, the beams are transversely polarized. Spin rotators located before and after the interaction regions are used to orient the proton spins longitudinally. The spin orientation of the beam can be varied from bunch to bunch and this pattern is systematically varied throughout the data collection period to help reduce and identify systematic errors associated with beam effects. The 
polarization of the beams are measured using a Hydrogen Jet Polarimeter~\cite{Zelenski:2005mz} and a pCarbon polarimeter~\cite{Alekseev:2003cn} which provide absolute and relative polarization measurements respectively. The protons are highly polarized 
when they are produced and the polarization is maintained through use of Siberian Snakes installed around the ring.

The Solenoidal Tracker at RHIC (STAR)~\cite{Ackermann:2002ad} is a multi-purpose, large acceptance detector located at the 6 o'clock position on the RHIC ring. The relevant sub-detectors used for this measurement are listed below. The beam line is encapsulated in azimuth by the Time Projection Chamber (TPC) which reconstructs the momenta of the charged particles. The TPC is 
followed by the Barrel (BEMC) and Endcap Electromagnetic Calorimeters (EEMC) which measure the energy of 
electromagnetically charged particles, photons and electrons in the case of STAR. In addition to these, various detectors are used for providing the local polarimetry and relative luminosity, such as the Beam Beam Counter (BBC), Vertex Position Detector (VPD) and Zero Degree Calorimeter (ZDC). The magnetic field is produced 
using a solenoidal magnet, which can generate varying magnetic fields from 0.25 to 0.5 Tesla.

\section{Analysis and Results}

Jets are showers of particles produced from the hadronization of quarks and gluons. Jets can be analysed at different levels - at the detector level in data and at the parton, particle and detector level in simulation. The detector level jets are reconstructed from TPC tracks and EMC towers, the particle level jets from stable final state particles produced during hadronization, and the parton level jets from the scattered partons produced in the hard collision as well as those from final and initial state radiation. At the parton level all the contributions from underlying events have been removed. STAR uses the FastJet \emph{$Anti-k_{T}$} algorithm~\cite{Cacciari:2008gp} to reconstruct jets. It is a sequential clustering algorithm that is less susceptible to effects from pile-up and underlying event contributions. It is also collinear and infrared safe to all orders. The radius parameter used for this analysis is R=0.5.

Events are recorded online if they satisfy the requirements of at least one of the active triggers during the run. The BEMC detector is divided into 18 overlapping jet patches, each consisting of 400 towers and spanning $\eta$ x $\phi$ = 1.0 x 1.0. The triggers used in this analysis are the jet patch triggers JP0, JP1 and JP2. Each of these triggers require a minimum transverse energy deposit in a patch of calorimeter towers. The minimum energy to be deposited are 5.4, 7.3 and 14.4 GeV respectively for JP0, JP1 and JP2 triggers. Dijet events are identified as those with at least two jets.  The two jets with the highest transverse momentum in the event are identified as the dijet pair and are subjected to the following cuts. An asymmetric $p_{T}$ cut is applied requiring one of the jets in the dijet pair to have $p_{T} > 8$ GeV and the other to have $p_{T} > 6$ GeV. One of the jets is required to be geometrically matched to one of the jet patches above the trigger threshold, to ensure that the reconstructed dijet caused the trigger for the event. The jets are required to be back-to-back $(\Delta\phi > 2.0)$ and to be within the detector's acceptance region $(-0.7<\eta_{det}<0.9)$. There is also a requirement that the calorimeters contribute less than 95\% of the total jet energy.

Once the dijets have been identified, they are spin-sorted and binned according to the invariant mass ($M_{inv}$). The $M_{inv}$ is calculated from the four-momentum of the dijet. The four-momenta of dijets are defined as the four-vector sum of the two jets:
\begin{equation}
p_{dijet} = p_1 + p_2
\end{equation} 
The $M_{inv}$ is defined as:
\begin{equation}
M_{inv} = \sqrt{E^2 - \vec{p}^2}
\end{equation}
For a single run, the $A_{LL}$ is calculated as:
\begin{equation}
A_{LL} = \left(\frac {1}{P_{Y}P_{B}}\right)\left(\frac{(N^{++}+N^{--})-R(N^{+-}+N^{-+})}{(N^{++}+N^{--})+R(N^{+-}+N^{-+})}\right)
\end{equation}
where $P_Y$ and $P_B$ are the polarizations of the two proton beams, R is the relative luminosity (the ratio of the luminosity for the aligned vs. anti-aligned bunches) and $N^{++}$, $N^{+-}$, $N^{-+}$ and $N^{--}$ are the spin-sorted yields. After calculating the $A_{LL}$ and its error for each run, the method of weighted averages is used to obtain the final values of $A_{LL}$ and its error for each bin.

\begin{figure}
\centering
\includegraphics[scale=0.577]{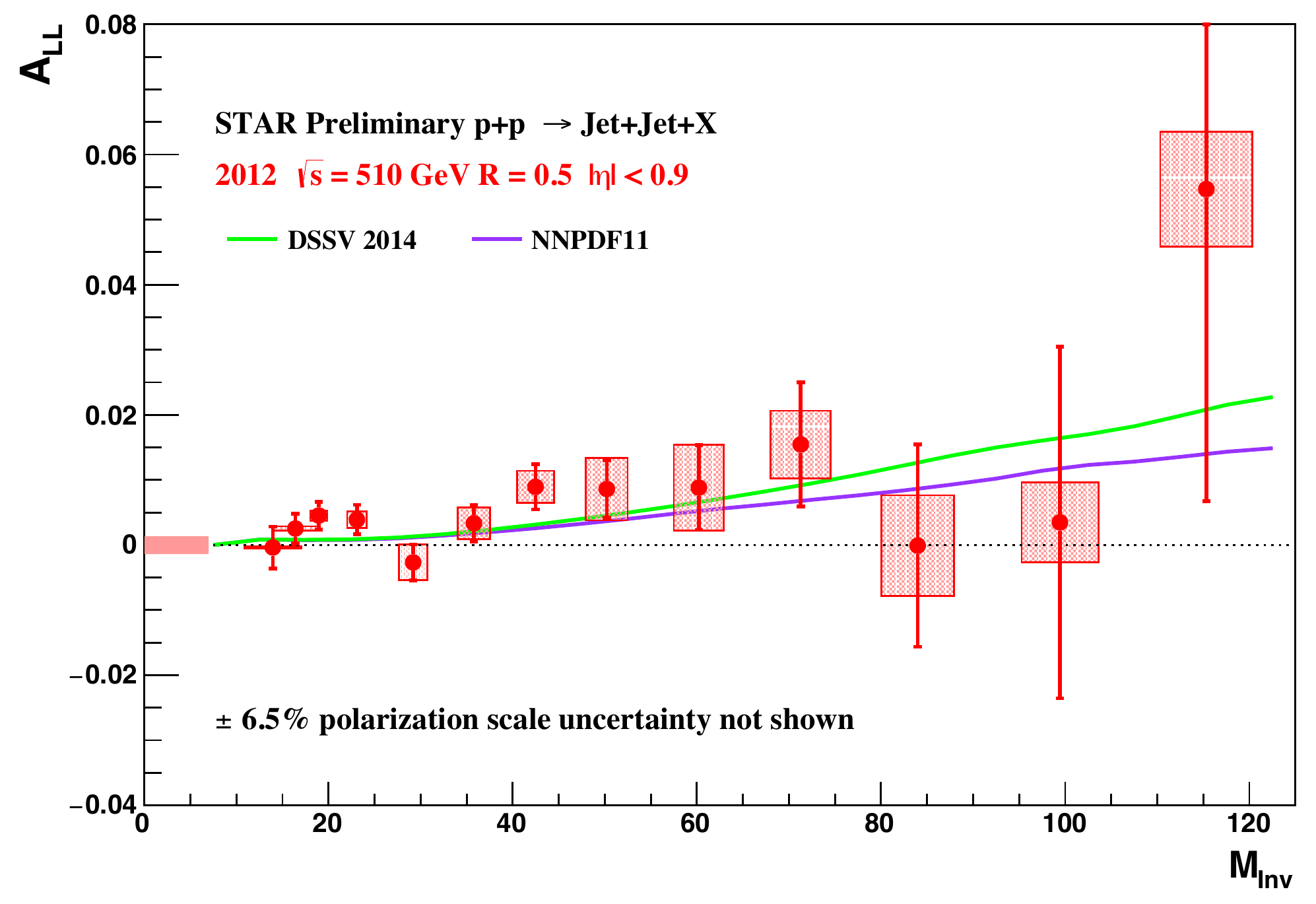}
\caption{Dijet $A_{LL}$ for $\sqrt s$ = 510 GeV as a function of $M_{inv}$ along with the theory curves for DSSV 2014~\cite{deFlorian:2014yva} and NNPDF1.1~\cite{Nocera:2014gqa}.}
\end{figure}

Figure 1 shows the preliminary results for the 2012 dijet $A_{LL}$ at 510 GeV for $|\eta|<0.9$ as a function of $M_{inv}$. The red points are data, with the errors bars for the statistical uncertainties. The red shaded boxes represent the systematic errors from jet energy scale and trigger bias. In addition to that there is a relative luminosity uncertainty of 4 x$10^{-4}$ shown as the red shaded horizontal band. The theory curves for DSSV 2014~\cite{deFlorian:2014yva} and NNPDF1.1~\cite{Nocera:2014gqa} are shown in green and purple, respectively. There is also a 6.5\% polarization scale uncertainty. It is clear that the data agrees well with both DSSV and NNDPDF.

Simulated \emph{pp} collisions are used to correct the reconstructed $M_{inv}$ for detector inefficiency and resolution effects. Monte Carlo (MC) events are generated with PYTHIA 6.426~\cite{Sjostrand:2006za} using the Perugia 0 tune~\cite{Skands:2010ak}, and are passed through detector simulations created in GEANT~\cite{Agostinelli:2002hh}. The shift in $M_{inv}$ is calculated by comparing the $M_{inv}$ distribution at the detector level to the 
parton level. The systematic error on the reconstructed $M_{inv}$ is due to the jet energy scale uncertainty. The jet energy scale systematic error includes uncertainties due to the TPC tracking efficiency as well as the BEMC calibration and hadronic energy deposition.

The trigger conditions used to select events may bias the measured jet distributions from the true jet distributions. For example, on average gluon jets are broader and have higher multiplicity than quark jets. Because the JP trigger patches have a finite acceptance, this may result in a change in the fractions of quark-gluon and gluon-gluon events sampled by the jet patch triggers. To estimate the effect of this bias theoretical parameterizations are used to weight MC events based on the $x$, $Q^2$ and the partonic flavor characteristics of the event. This event re-weighting is used to create $A_{LL}$ predictions at the detector and parton levels. For a given set of polarized structure functions, after evaluating the $A_{LL}$ at both the detector and parton levels, the uncertainty is calculated as the difference between these parton and detector $A_{LL}$. This procedure is repeated for various models and the uncertainity is taken from the model with the largest difference, which was the LSS10p ~\cite{Leader:2010rb}model in this case.

\begin{figure}
\centering
\includegraphics[scale=0.577]{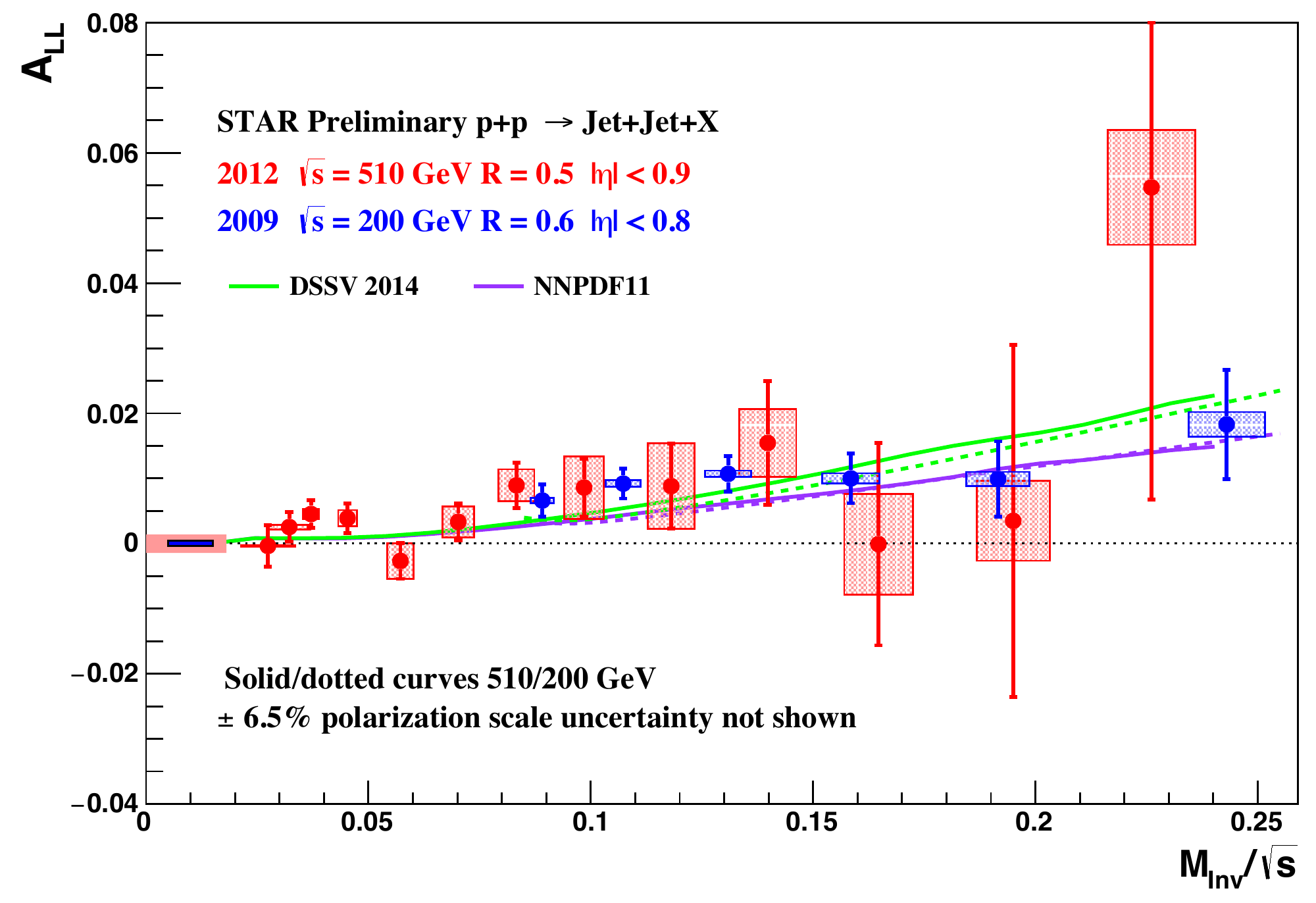}
\caption{Dijet $A_{LL}$ as a function of $\frac{M_{inv}}{\sqrt s}$ for $\sqrt s$ = 200 GeV~\cite{Aschenauer:2016our} and 510 GeV along with the theory curves for DDSV 2014~\cite{deFlorian:2014yva} and NNPDF1.1~\cite{Nocera:2014gqa}.}
\end{figure}

Figure 2 compares the dijet $A_{LL}$ results for 510 GeV and 200 GeV~\cite{Aschenauer:2016our} from 2012 and 2009 respectively, as a function of $x_{T} = \frac{M_{inv}}{\sqrt s}$ so they can be shown on the same scale. The blue points with the error bars are the 200 GeV data with statistical uncertainties, and the blue shaded boxes represent the systematics. For $x_{T} > 0.1$, the results from 200 GeV and 510 GeV are consistent and the higher $\sqrt s$ of 510 GeV clearly extend the measurements to lower $x_{T}$. It should also be noted that $x_{T}$ scales with $x$ and therefore sampling lower $x_{T}$ dijets accesses low $x$ partons.

\section{Conclusions}

The highly polarized proton beams at RHIC have facilitated a robust spin program at STAR. The wide acceptance of the detector is well suited for jet reconstruction. The STAR 
inclusive jet measurements at $\sqrt s=200$ GeV have provided the first evidence of a significant polarized gluon distribution for $x > 0.05$. By extending these measurements to a higher $\sqrt s$, it is 
possible to constrain the $x < 0.05$ region, and dijet observables allow for reconstruction of the partonic kinematics at leading order. The preliminary results from the first dijet $A_{LL}$ measurement at $\sqrt s=510$ GeV from the 2012 RHIC run have been shown, and they agree well with the previous measurements at 200 GeV and that of theoretical NLO calculations. In 2013, STAR collected thrice more data, of longitudinally polarized proton collisions, which will provide even tighter constraints on this measurement in the future.

\end{document}